# Robust Ti$^{4+}$ states in SrTiO$_3$ layers of La$_{0.6}$Sr$_{0.4}$MnO$_3$/SrTiO$_3$/La$_{0.6}$Sr$_{0.4}$MnO$_3$ junctions

H. Kumigashira,[a)] A. Chikamatsu, R. Hashimoto, and M. Oshima

*Department of Applied Chemistry, The University of Tokyo, Bunkyo-ku, Tokyo 113-8656, Japan*

T. Ohnishi and M. Lippmaa [b)]

*Institute for Solid State Physics, The University of Tokyo, Kashiwa 277-8581, Japan*

H. Wadati and A. Fujimori

*Department of Physics and Department of Complexity Science and Engineering, The University of Tokyo, Bunkyo-ku, Tokyo 113-0033, Japan*

K. Ono

*Institute of Materials Structure Science, KEK, Tsukuba 305-0801, Japan*

M. Kawasaki [b)]

*Institute for Materials Research, Tohoku University, Sendai 980-8577, Japan*

H. Koinuma [b)]

*National Institute for Materials Science, Tsukuba 305-0047, Japan*

**Abstract**

We have investigated the interfacial electronic structure of $La_{0.6}Sr_{0.4}MnO_3$ (LSMO)/ $SrTiO_3$ (STO)/ LSMO heterojunctions utilizing the elemental selectivity of photoemission spectroscopy. The Ti 2*p* core-level spectra clearly show $Ti^{4+}$ states and do not exhibit any indication of $Ti^{3+}$ states in $TiO_2$ layers irrespective of a different kind of adjacent atomic layer with different chemical carrier concentration. This result indicates that the Ti ions in the $TiO_2$ atomic layers preserve their tetravalent states even in the vicinity of the valence-mismatched interface between LSMO and STO, reflecting chemical stability of the $Ti^{4+}$ states.

a) Electronic mail: kumigashira@sr.t.u-tokyo.ac.jp
b) also at Combinatorial Materials Exploration and Technology (COMET), Tsukuba 305-0044, Japan

Spin tunnel junctions employing half metals as electrodes have attracted considerable attention because of their potential applications to magneto-electronic devices. The hole-doped perovskite manganite, $La_{0.6}Sr_{0.4}MnO_3$ (LSMO), is one of the most promising materials for such applications [1]. In principle, perfect spin polarization of conduction carriers in LSMO should make the tunneling magnetoresistance (TMR) ratio huge in heterostructures [2,3]. However, in actual spin-tunnel junction devices, such as the LSMO/$SrTiO_3$ (STO)/LSMO trilayers, in which STO is used as a tunnel barrier, the performance is far worse than what would be expected from the high spin polarization and high Curie temperature of LSMO [4-6]; The TMR ratio has been much smaller than that would be expected from the half-metallic nature of LSMO and the TMR response quickly diminishes at elevated temperatures, even far below the Curie temperature ($T_C$) of LSMO. A possible origin of the degraded TMR is the formation of a kind of dead layer at the LSMO/STO valence-mismatched interfaces [6]. Since the magnetic and electronic properties of oxides can be easily affected by interface effects, such as spin exchange interactions [7], charge transfer [8,9], and epitaxial strain [10], it is critically important to understand the electronic structure at oxide heterointerfaces. However, the lack of information on the electronic structure of the interface layers, especially the valency of transition metals in the constituent layers, does not allow us to fabricate heterojunctions with predetermined properties.

Photoemission spectroscopy (PES) is a powerful technique for studying this issue, since it provides detailed information on the electronic structure at interfaces.

Direct observation of the element-specific electronic structures in constituent layers is a key to understanding the interfacial electronic structure, which dominates the TMR characteristics of junction devices.  In this letter, we report on *in situ* Ti 2*p* core-level PES characterization of tunnel barrier STO layers which are sandwiched between LSMO layers to investigate the interfacial electronic structure of LSMO/STO/LSMO junctions.  In particular, we focus on possible reduction of Ti ions in the barrier layers. We present spectral evidence that the Ti ions preserve their tetravalent states even at the interface between LSMO and STO layers owing to their chemical stability.

The STO/LSMO and LSMO/STO/LSMO multilayers as well as LSMO thin films were fabricated onto atomically flat $TiO_2$-terminated STO (001) substrates in a laser MBE chamber connected to a synchrotron radiation photoemission system at BL2C of the Photon Factory, KEK [11, 12].  Schematic side views of the fabricated multilayers are shown in the bottom panel of Fig.1.  Sintered LSMO and STO pellets were used as ablation targets.  A Nd:YAG laser was used for ablation in its frequency-tripled mode ($\lambda = 355$ nm) at a repetition rate of 1 Hz.  During deposition, the substrate temperature was kept at 950°C and the oxygen pressure was $1\times10^{-4}$ Torr.  The STO layer thickness was varied from 2 to 5 ML, while the thicknesses of the bottom and top LSMO layers were fixed at 50 ML and 5 ML, respectively.  The thickness of each layer was controlled on an atomic scale by monitoring the intensity oscillation of the specular spot in reflection high-energy electron diffraction (RHEED) during growth. These samples were subsequently annealed at 400°C and atmospheric pressure of oxygen to remove oxygen vacancies.  After cooling down below 100°C, the samples were transferred into the photoemission chamber under a vacuum of $10^{-10}$ Torr. The in-vacuum transfer is necessary to avoid defect formation and/or extrinsic chemical reactions caused by surface cleaning procedures done in earlier *ex situ* PES measurements [11, 12]. The PES spectra were taken using a Scienta SES-100 electron energy analyzer with a total energy resolution of 150 meV in the 600-800 eV photon energy range. The surface stoichiometry of the samples was carefully characterized by analyzing the relative intensity of relevant core levels, and we confirmed that the composition of the samples was the same as the ceramic targets.

The surface morphology of the multilayers was analyzed by *ex situ* atomic force microscopy (AFM). Step-and-terrace structures were clearly observed as shown in Fig. 1. The atomically flat surface of the films and multilayers suggests that the interface structure is also controlled on an atomic level. We have also confirmed the formation of a chemically abrupt interface between the STO and LSMO layers by comparing the relative intensities of core levels as a function of the STO cover layer thickness with a simulated photoelectron attenuation function [13]. In addition, detailed analysis of the terminating layer of these multilayers by measuring the angular dependence of core-level spectra revealed that these multilayers are terminated by B-site layers, namely the $TiO_2$ and $MnO_2$ layers for STO/LSMO and LSMO/STO/LSMO multilayers, respectively [12].

Figure 2 shows the Ti $2p$ core-level spectra of the STO/LSMO and LSMO/STO/LSMO multilayers, together with a spectrum of a STO single crystal. The binding energies of $Ti^{4+}$ peaks appeared at about 459.2 eV in STO and 458.0 eV in STO/LSMO multilayers, respectively, which we set the zero for relative binding energies. The slight energy shift of 1.2 eV in STO/LSMO multilayers originates from band bending between the insulating STO and metallic LSMO layers, rather than a chemical shift, because all core levels are shifted by the same amount. The observed Ti 2$p$ core level is indicative of typical $Ti^{4+}$ states [14-16]. It is well established that $Ti^{3+}$ components emerge at the lower binding energy side with energy separation of about 1.5 eV from $Ti^{4+}$ states [14,15]. Assuming the reduction of Ti states due to charge transfer between STO and LSMO layers, significant $Ti^{3+}$ signals should be expected to appear at the lower binding energy side of the $Ti^{4+}$ peaks. However, we do not observe any indication of $Ti^{3+}$ states within the experimental errors. The absence of $Ti^{3+}$ states is clearly seen by comparing the Ti 2$p$ core level of overlayers with that of STO single crystals ($Ti^{4+}$ states), as shown in the inset of Fig. 2 (a). The lack of any significant changes with increasing STO layer thickness indicates that Ti ions are present only in the $Ti^{4+}$ states in all multilayers. These results show that there is no charge transfer across the interface from LSMO to STO unlike the LSMO/LSFO heterointerface, where strong charge redistribution has been observed [13].

There is an important interface problem in perovskite oxide superlattices, represented by the formula $ABO_3$/A'B'$O_3$/$ABO_3$ [6]. Since the perovskite oxides consist of alternating AO (A-site plane) and $BO_2$ (B-site plane) atomic layers, stacked along the $c$-axis direction, the $ABO_3$/A'B'$O_3$ and A'B'$O_3$/$ABO_3$ interface structures are not identical; the former interface has the $BO_2$/AO//B'$O_2$/A'O layer sequence, whereas the latter has the B'$O_2$/A'O//$BO_2$/AO structure. Thus, in the case of STO/LSMO, the $TiO_2$ planes in STO overlayers are surrounded by SrO layer, preserving the $Ti^{4+}$ state in the vicinity of the interface. In contrast, we get a valence-mismatched interface when the terminating layers at the interface are switched, as in a -$MnO_2$-$La_{0.6}Sr_{0.4}O$-$TiO_2$-SrO- stacking sequence, where the $La_{0.6}Sr_{0.4}O$ atomic layer is expected to act as an "electron-donor" layer for the $TiO_2$ layer [5,6]. In order to investigate the charge transfer at such an interface, we have measured the Ti 2*p* core levels for STO layers sandwiched between LSMO layers where the top $TiO_2$ plane adjacent to a LSMO overlayer faces to a $La_{0.6}Sr_{0.4}O$ plane. The results are shown in Fig. 2 (b). Even at a valence-mismatched interface, the Ti 2*p* core-level spectra do not show any indication of the presence of $Ti^{3+}$ states within the experimental errors, just like in STO/LSMO multilayers where the interface consists of a stacking sequence of -$TiO_2$-SrO-$MnO_2$-$La_{0.6}Sr_{0.4}O$-. This result clearly shows that the chemical states of $Ti^{4+}$ ions are preserved even in $TiO_2$ atomic planes adjacent to the "electron donor" $La_{0.6}Sr_{0.4}O$ atomic layers.

The observed robust chemical states of Ti ions are in sharp contrast to the electronic structure of $La_{0.6}Sr_{0.4}FeO_3$ (LSFO)/LSMO interfaces, where interfacial charge transfer occurs through the heterointerface owing to the different 3*d* energy levels of constituent transition metals [13]. The charge transfer between the constituent transition metals seems to be common phenomena at heterointerfaces based on perovskite oxides such as the $LaFeO_3$/$LaCrO_3$ [8] and $CaRuO_3$/$CaMnO_3$ systems [9]. The unique behavior in the present system can be explained by considering the electronic structures of titanium, manganese, and iron as illustrated in Fig. 3. In perovskite oxides, the 3$d$ orbitals of transition-metal ions are known to split into $t_{2g\uparrow}$, $e_{g\uparrow}$, $t_{2g\downarrow}$, and $e_{g\downarrow}$ levels (in the order of increasing energy), owing to the octahedral crystal field and strong Hund's coupling. At a heterointerface, the excess carriers in an AO layer dominate the filling of the 3$d$ states of transition metals in the adjacent BO$_2$ layers, since the AO layer is shared between the BO$_2$ and B'O$_2$ layers. The La$_{0.6}$Sr$_{0.4}$O atomic layer acts as an electron-donor layer for an adjacent BO$_2$ layer, while the SrO layer does not.

By assuming a hypothetical single BO$_2$ layer, the electronic configurations are $t^0_{2g\uparrow}$ for Ti ions in a TiO$_2$ layer, $t^3_{2g\uparrow}e^0_{g\uparrow}$ for Mn ions in a MnO$_2$ layer, and $t^3_{2g\uparrow}e^1_{g\uparrow}$ for Fe ions in a FeO$_2$ layer. When the electron-donor La$_{0.6}$Sr$_{0.4}$O layer, which donates 0.6 electrons per lattice site to the transition metal oxide layer, is shared by MnO$_2$ and FeO$_2$ layers, the excess electrons are expected to be accumulated in the $e_{g\uparrow}$ states of Fe, owing to the energy difference of the partially occupied $e_{g\uparrow}$ states in Fe and Mn ions as illustrated in Fig. 3 (b) [13]. As a result, the effective hole concentration increases in the first few LSMO unit cell layers close to the LSFO interface. On the contrary, the lowest energy level of 3$d$ states in STO are empty $t_{2g\uparrow}$ states which are located above the Fermi level ($E_F$) with energy separation of about 1.0 eV from Mn 3$d$ $e_{g\uparrow}$ states [16]. Owing to the energy difference of the 3$d$ states in transition metal ions, it is clear that the excess electrons in the La$_{0.6}$Sr$_{0.4}$O atomic layer are accommodated in the Mn ions of the LSMO/STO (-SrO-TiO$_2$-La$_{0.6}$Sr$_{0.4}$O-MnO$_2$-) interface owing to different energy level between LSMO and STO. As a consequence, the Ti$^{4+}$ states are preserved even in the valence-mismatched interfaces.

On the other hand, for the STO /LSMO (-TiO$_2$-SrO-MnO$_2$-La$_{0.6}$Sr$_{0.4}$O-) interface, the roubst Ti$^{4+}$ states result in an effective hole concentration increase in the first few LSMO unit cell layers close to the STO interface.  A preliminary Mn 2$p$-3$d$ resonant PES study on STO/LSMO multilayers has revealed the occurrence of execss hole-doping in the LSMO layer, as expected [17].  These results suggest the importance of charge distribution, which is dominated by the energies of the 3$d$ levels of constituent transition metals in a heterostructure.  A junction therefore needs to be designed by tuning the local carrier concentration independently for each atomic layer in the vicinity of the interface.  If such atomic-layer level design is not performed, the conductivity and spin polarization would be significantly degraded at the tunnel barrier interfaces, which is the most critical part of TMR devices.

This work was done under Project No. 02S2-002 at the Institute of Materials Structure Science at KEK. This work was supported by a Grant-in-Aid for Scientific Research (A16204024 and S17101004) from the Japan Society for the Promotion of Science.

**Figure Captions**

Fig. 1: AFM images (upper panel) and schematic side views (bottom panel) of (a) an LSMO film, (b) STO/ LSMO bilayer with a STO overlayer thichness of $n$ = 3 ML, and (c) LSMO/ STO/ LSMO multilayer where a 3-ML STO layer is sandwitched between the top LSMO (5ML) and bottom LSMO (50ML) layers.  The scan area of the AFM images is 500 x 500 nm$^2$.

Fig. 2: (color online) Ti 2$p$ core-level photoemission spectra of SrTiO$_3$ layers with different layer thicknesses for (a) STO/LSMO and (b) LSMO/STO/LSMO multilayers. Ti 2$p$ spectra of SrTiO$_3$ single crystal are also presented for comparison.  The insets show the spectra around Ti$^{4+}$ states on an expanded energy scale.  Note that the Ti$^{3+}$ states, which are expected to appear at the lower binding energy side of Ti$^{4+}$ states with an energy separation of about -1.5 eV [14], seem to be absent within the experimental error margins.

Fig. 3: A schematic diagram of (a) the electronic structure of Ti and Mn ions at LSMO/STO heterointerfaces, and (b) that of Mn and Fe ions at LSFO/LSMO heterointerfaces.  The energy positions of 3$d$ levels for each ion were taken from the previous PES experiments [13,16].

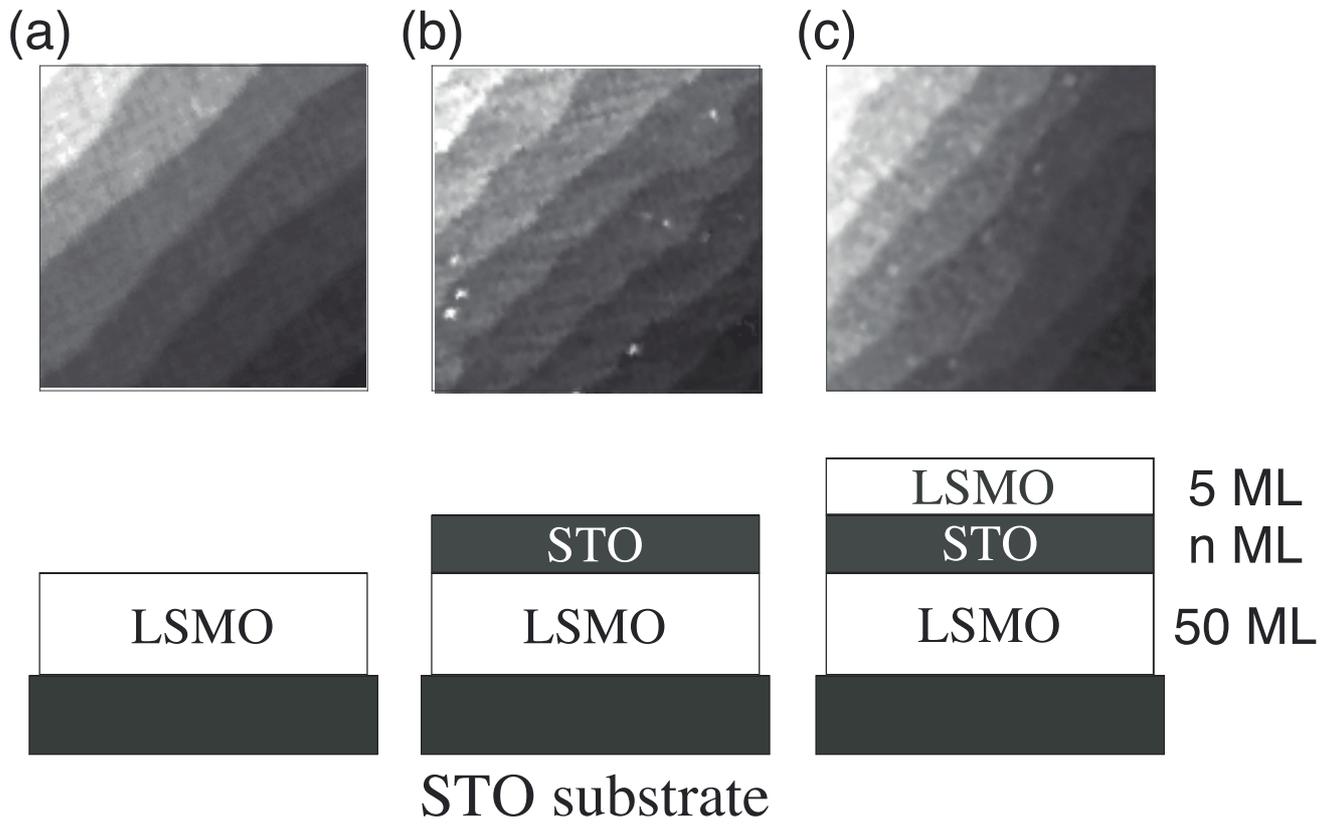

Figure 1. H. Kumigashira *et al.*, Appl. Phys. Lett.

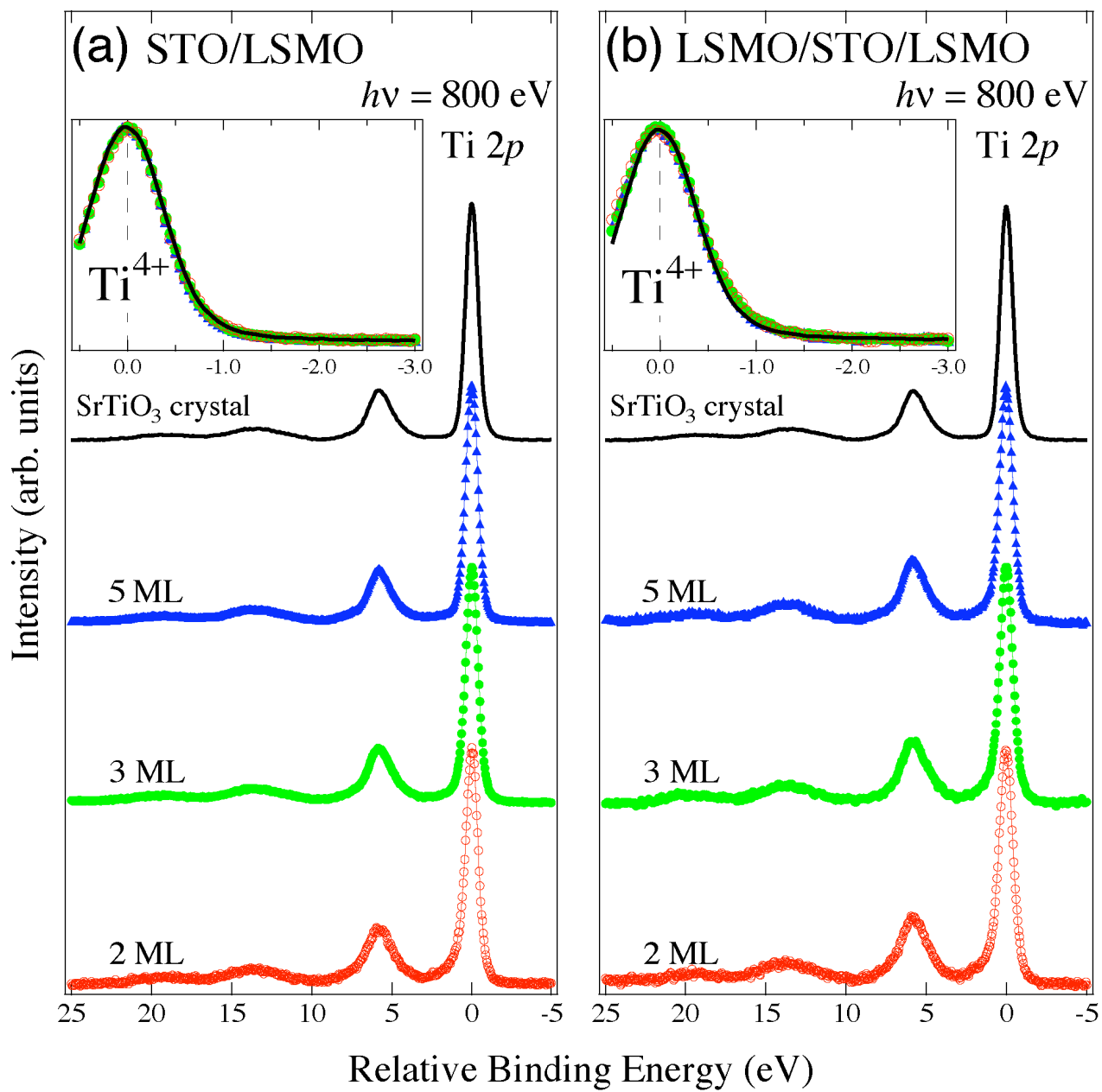

Figure 2. H. Kumigashira *et al.*, Appl. Phys. Lett.

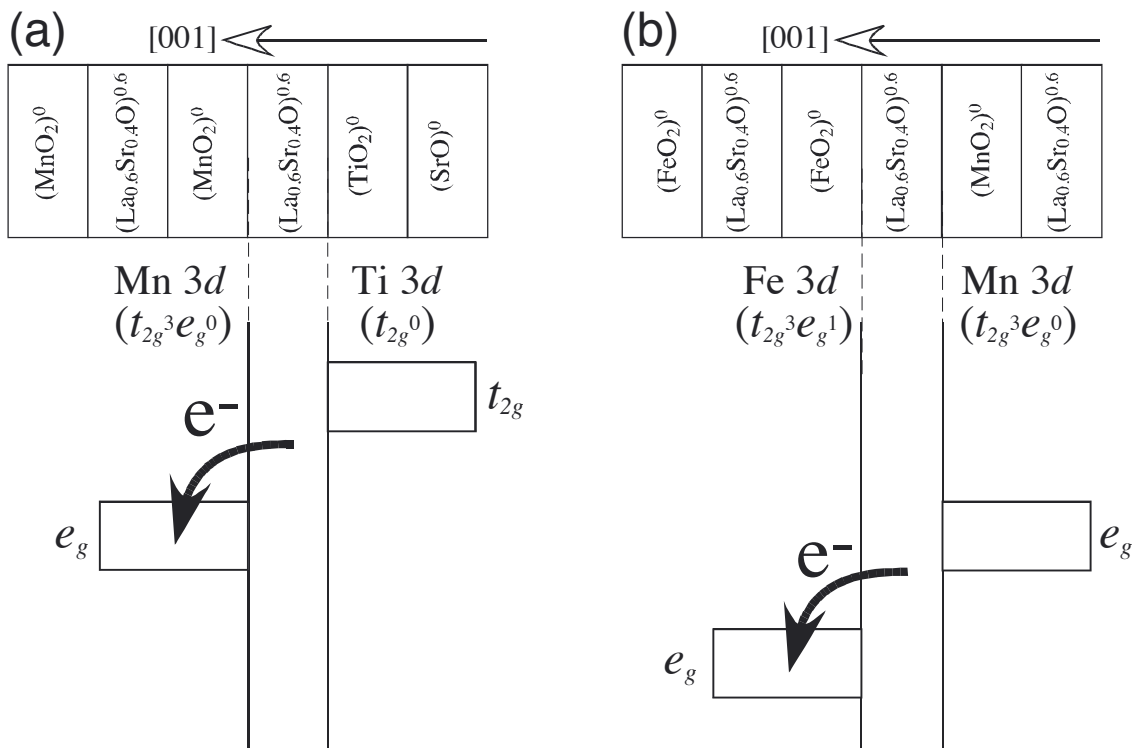

Figure 3. H. Kumigashira *et al.*, Appl. Phys. Lett.